\documentclass[conference]{IEEEtran}
\IEEEoverridecommandlockouts
\usepackage{cite}
\usepackage{amsmath,amssymb,amsfonts}
\usepackage{algorithmic}
\usepackage{graphicx}
\usepackage{booktabs}
\usepackage{textcomp}
\usepackage{xcolor}
\usepackage{subfig}
\usepackage[hidelinks]{hyperref}

\def\BibTeX{{\rm B\kern-.05em{\sc i\kern-.025em b}\kern-.08em
    T\kern-.1667em\lower.7ex\hbox{E}\kern-.125emX}}
\begin{document}

\title{\Huge{Exploring the Interplay of Interference and Queues in Unlicensed Spectrum Bands for UAV Networks}
}

\author{

\IEEEauthorblockN{Masoud Ghazikor\IEEEauthorrefmark{1}, Keenan Roach\IEEEauthorrefmark{2}, Kenny Cheung\IEEEauthorrefmark{2}, Morteza Hashemi\IEEEauthorrefmark{1}}
        \IEEEauthorblockA{
        \IEEEauthorrefmark{1}Department of Electrical Engineering and Computer Science, University of Kansas \\
        \IEEEauthorrefmark{2}Universities Space Research Association (USRA)
        }
}

\maketitle

\begin{abstract}
In this paper, we present an analytical framework to explore the interplay of signal interference and transmission queue management, and their impacts on the performance of unmanned aerial vehicles (UAVs) when operating in the unlicensed spectrum bands. In particular, we develop a comprehensive framework to investigate the impact of other interference links on the UAV as it communicates with the ground users. To this end, we provide closed-form expressions for packet drop probabilities in the queue due to buffer overflow or large queuing delay, which are expressed in terms of a transmission policy as a function of the channel fading threshold $\beta$. The overall packet loss caused either by interference signals or queuing packet drop is obtained, which, in turn, yields in obtaining the expected throughput performance. Through extensive numerical results, we investigate the impact of the channel fading threshold $\beta$, which plays an important role in balancing the trade-offs between packet loss due to queue drop or transmission error due to large interference levels.

\end{abstract}


\section{Introduction}
An unmanned aerial vehicle (UAV), commonly known as a drone, is an aerial vehicle that can be controlled remotely from the ground control station or be preprogrammed to fly autonomously. UAVs have shown great potential for a variety of applications in wireless communications due to their excellent communication links with the ground, easy deployment, and ability to perform different tasks \cite{Shen-2020-Multi}. To this end, UAV communication needs to provide high reliability and low latency to ensure the safe functioning of UAVs~\cite{Maolin-2022-Analysis, Yan-2019-Comprehensive}. 

In general, UAV communications can utilize both licensed and unlicensed spectrum. Unlicensed spectrum, being shared among various users, comes with lighter regulations, making users more susceptible to interference from others. This susceptibility is the main contributor to throughput performance degradation. Two other significant factors affecting throughput performance involve the transmitter queue: firstly, dropping packets when the waiting time exceeds a threshold due to bad channel conditions, such as high interference levels; and secondly, the risk of buffer overflow arising from the finite buffer size, potentially leading to performance degradation.

Prior works mostly focused on energy efficiency, optimal positioning, and coverage or outage analysis of the UAVs \cite{Zeng-2017-Energy}\cite{Hosseinalipour-2019-Interference}\cite{Liu-2018-Performance}, but there are a few works that considered the impact of multi-user interference and transmitter queue management on UAVs communication and their relationship with each other which affects the overall throughput. In \cite{Kim-2018-Outage}, ground-to-air (G2A), air-to-ground (A2G), ground-to-ground (G2G), and air-to-air (A2A) channels are modeled by Rayleigh or Rician distributions, and the outage probability is calculated based on one interference link only. In \cite{Khuwaja-2019-Optimum}, ground users are served by a UAV in the presence of interferer UAVs, while other users are not considered as interferer nodes.
In \cite{Shi-2021-Coverage}, the channel model is considered as Nakagami fading channel constrained in mmWave, and the path loss exponent is assumed constant for both LoS and NLoS. In \cite{Abualhaol-2006-Outage}, the outage probability is expressed in terms of the signal-to-noise ratio (SNR), without considering the impact of interference. In \cite{Azari-2018-Ultra}, only the A2G channel is modeled by the Rician distribution to calculate the outage probability without impact of interference again.

\begin{figure}
    \centering
   \includegraphics[width=0.85\linewidth]{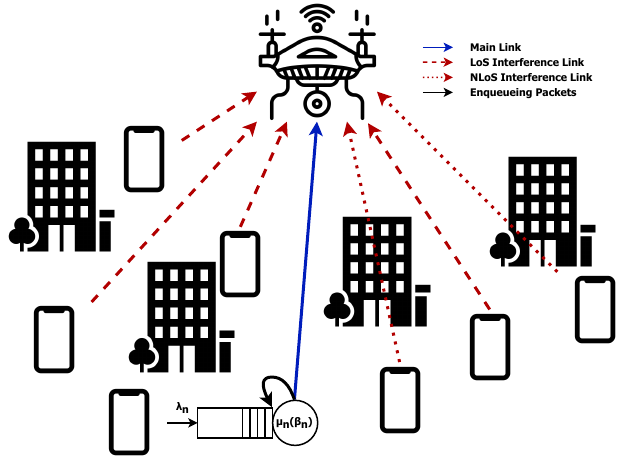}
    \caption{System model consists of a source node, interference nodes, and UAV with two types of LoS and NLoS channels.}
    \label{Fig. 1}
\end{figure}

As mentioned, there are many prior works that consider the UAV communication models, but the problem of exploring the interplay between UAV interference and queue management in unlicensed spectrum bands is not fully investigated, and prior system models and solutions do not completely capture all the factors that impact the UAVs performance. In this paper, we express the general formula to calculate the expected throughput that consists of (i) the probability of transmission error by considering the SINR as the contributing factor for packet drops, (ii) the probability of dropping the packet due to exceeding the maximum queuing delay threshold, and (iii) the probability of buffer overflow that depends on the offered load. Our system model captures both LoS and NLoS links between any pairs of transmitter and receiver, including both ground and aerial users. In summary, the main contribution of this paper is summarized as follows:

\begin{itemize}
    \item We consider a comprehensive model to calculate the expected throughput of UAVs in unlicensed spectrum bands. To this end, our solution takes into account two steps where packets can be dropped. The first step is analyzing the packet drop in the transmitter queue (probability of buffer overflow and large queuing delay) and the second step is considering the packet drop after transmitting the packet (probability of transmission error). 
    \item We analyze the probability of transmission error between the source node and UAV in the presence of interferer nodes where channels could be Rayleigh (NLoS) or Rician (LoS) based on UAV's height and LoS probability. 
    \item We investigate the channel fading threshold for both Rayleigh and Rician channels and implement the Jacobi best-response algorithm to analyze the behavior of the channel fading threshold versus expected throughput.
\end{itemize}
It should be noted that the analytical framework presented in this paper is motivated by the work in \cite{Guan-2016-ToTransmit}, while (i) considering both ground-level and UAV users as opposed to ground users only in \cite{Guan-2016-ToTransmit}, and (ii) considering transmit queues with finite buffer, which could result in a buffer overflow and performance degradation for data-intensive UAV applications. 

The rest of this paper is organized as follows. In Section \ref{system model}, we present the system model, LoS probability, path loss, and channel model. Section \ref{queueing analysis} provides a detailed analysis of queueing delay and buffer overflow. In Section \ref{impacts of interference}, we investigate the impact of interference on the main link and express the probability of transmission error due to low SINR. Section \ref{throughput performance analysis} presents the overall packet loss, and the expected throughput. In Section \ref{numerical results}, we provide an experimental evaluation of our model. Finally, Section \ref{conclusion} concludes the paper.

\section{System Model} \label{system model}
The system model is shown in Fig. \ref{Fig. 1} where the source node establishes a main link with UAV and other nodes operating as interferer nodes. The source node can send packets toward UAV or enqueue them according to the channel condition.

Next, we define the LoS probability ($P_{LoS}$) to determine the different types of channels and obtain the path loss according to $P_{LoS}$. Then, we analyze two channel models (Rayleigh and Rician) for unlicensed UAV operations.

\subsection{Line-of-Sight Probability}
Consider a source node and UAV as transmitter and receiver, respectively, also some interference nodes. We calculate $P_{LoS}$ to find channel's type (LoS or NLoS) among nodes. Initially, we need to determine the vertical distance $d_i^{V} = \sqrt{(z_{i}-z_{u})^2}$ and the horizontal distance $d_i^{H} = \sqrt{(x_{i}-x_{u})^2 + (y_{i}-y_{u})^2}$ between the transmitter and receiver ($d_n^{V},d_n^{H}$) and between the interference nodes and  receiver ($d_{\boldsymbol{m}}^{V},d_{\boldsymbol{m}}^{H}$). Then, the elevation angle can be expressed as \cite{Kim-2018-Outage}:
\begin{align}
\theta_i = \arctan{(\frac{d_i^V}{d_i^H})} \quad \forall{i} = \left\{n,\boldsymbol{m}\right\}. 
\end{align}
Based on the elevation angle, the $P_{LoS}$ is given by \cite{Azari-2018-Ultra}: 
\begin{align} \label{plos}
P_{LoS}(\theta_i) = \frac{1}{1+a_1e^{-b_1\theta_i}}, 
\end{align}
where $a_1$ and $b_1$ are environmental parameters. $P_{LoS}(\theta_i)$ in (\ref{plos}) is used mainly for the G2A and A2G channels but for the G2G channel $P_{LoS}(0) \rightarrow 0$ and the A2A channel $P_{LoS}(\frac{\pi}{2}) \rightarrow 1$.

\subsection{Path Loss Model}
Let $N$ be a set of communication sessions using the same spectrum band which is partitioned into a set of $F$ frequency channels and $n \in N$ be each session between source and destination. The relationship between the transmission power $P_t$ and the received power $P_r$ is given by
$P_r = P_t |h_n^f|^2,$
where $h_n^f$ is the channel gain $f \in F$, defined as $h_n^f = \Tilde{h}_n^f \hat{h}_n^f$ \cite{Guan-2016-ToTransmit}, where $\Tilde{h}_n^f$ is the channel fading coefficient and $\hat{h}_n^f$ is the square root of single-slope path loss which is defined as \cite{Ren-2011-Modelling}: 
\begin{align}
\hat{h}_n^f = \sqrt{K(\frac{d_0}{d_n})^{\alpha(\theta_i)}} \quad \text{if} \enskip d_n \ge d_0, 
\end{align}
in which $d_0$ is a reference distance,   
$d_n$ is the distance between the transmitter and receiver, and $K = \frac{\lambda^2}{16\pi^2d_0^2}$ is a constant factor.
Also, $\alpha(\theta_i)$ denotes the path loss exponent which is given by $\alpha(\theta_i) = \alpha_{\frac{\pi}{2}}P_{LoS}(\theta_i) + \alpha_0(1-P_{LoS}(\theta_i))$ \cite{Azari-2018-Ultra}. Finally, $\hat{h}_n^f$ can be expressed as: 
\begin{align}
\hat{h}_n^f = \frac{c}{4\pi f}\sqrt{\frac{d_0^{\alpha(\theta_i)-2}}{d_n^{\alpha(\theta_i)}}}. 
\end{align}

\subsection{Channel Model}
Consider a block fading channel model, where $\Tilde{h}_n^f$ could be Rician (Rice) or Rayleigh (Ray) distributions based on being LoS or NLoS channels, respectively. Let us initially focus on the Rician channel for which the probability density function (PDF) is given by: 
\begin{align}
Pb(\Tilde{h}_{n}^f = x) = xe^{-\frac{x^2+b^2}{2}}I_0(xb), 
\end{align}
where $I_0$ is the modified Bessel function of the first kind with order zero, and $b = \sqrt{2K(\theta_i)}$ is defined based on the Rician shape parameter $K(\theta_i)$ which can be expressed as~\cite{Kim-2018-Outage}: 
\begin{align}
 \quad K(\theta_i) = k_0e^{\frac{2}{\pi}\ln(\frac{k_{\frac{\pi}{2}}}{k_0})\theta_i}, 
\end{align}
By adopting a similar approach as in~\cite{Guan-2016-ToTransmit}, consider a channel threshold policy where $\beta_n > 0$ is defined as a channel fading threshold. Based on $\beta_n$, the source node $n \in N$ decides to transmit a packet toward its destination over the best frequency channel $f^* = \underset{f \in F}{\arg\max} \ \ 
\Tilde{h}_n^f \hat{h}_n^f$ if $\Tilde{h}_n^{f^*} \ge \beta_n$, otherwise enqueue the packet in its buffer. Then, the cumulative distribution function (CDF) of the Rician distribution can be defined as: 
\begin{align}
\begin{aligned}
Pb(\Tilde{h}_{n}^f < \beta_n^{Rice}) = \int_{0}^{\beta_n^{Rice}} xe^{-\frac{x^2+b^2}{2}}I_0(xb) \,dx
\\
= 1-Q_1(b,\beta_n^{Rice}), 
\end{aligned}
\end{align}
where $Q_1$ is the first-order Marcum Q-function \cite{You-2019-Trajectory}. Then, the probability of transmitting a packet during a time slot by the source node can be calculated as $\mu_n(\beta_n) = 1-Pb(\Tilde{h}_n^{f^*}<\beta_n)$. Thus, by assuming $|F|$ frequency bands,  we obtain: 
\begin{align}
\mu_n(\beta_n^{Rice}) = 1-(1-Q_1(b,\beta_n^{Rice}) )^{|F|}. 
\end{align}

A similar approach can be used for the Rayleigh channel which the PDF of Rayleigh distribution is expressed as: 
\begin{align}
Pb(\Tilde{h}_{n}^f = x) = \frac{2x}{\Omega}e^{-\frac{x^2}{\Omega}}, 
\end{align}
where $\Omega$ represents the Rayleigh fading factor. Then, the CDF of Rayleigh distribution can be defined as: 
\begin{align*}
\begin{aligned}
Pb(\Tilde{h}_{n}^f < \beta_n^{Ray}) = \int_{0}^{\beta_n^{Ray}} \frac{2x}{\Omega}e^{-\frac{x^2}{\Omega}} \,dx= 1-e^{-\frac{(\beta_n^{Ray})^2}{\Omega}}. 
\end{aligned}
\end{align*}
Therefore, the probability of transmitting a packet over the Rayleigh channel is obtained as follows: 
\begin{align}
\mu_n(\beta_n^{Ray}) = 1-(1-e^{-\frac{(\beta_n^{Ray})^2}{\Omega}})^{|F|}. 
\end{align}

\section{Queuing Analysis} \label{queueing analysis}
Similar to the methodology presented in \cite{Guan-2016-ToTransmit} by using an exponential distribution, the PDF of $v_n$ for both channels that represents the number of time slots required by the source node to transmit a packet to its destination can be approximated as:
\begin{align} \label{approximated pdf}
Pb(v_n=k) \approx \mu_n(\beta_n)e^{-\mu_n(\beta_n)k}. 
\end{align}

By applying Eq. (\ref{approximated pdf}), the probability of dropping a packet $P_n^{dly} (\beta_n)$ in the queue due to exceeding the maximum queuing delay threshold $T_n^{th}$, can be expressed as: 
\begin{align}
P_n^{dly} (\beta_n) \triangleq Pb(T_n > T_n^{th}) = e^{-(\frac{\mu_n(\beta_n)} {T_{slt}} - \lambda_n) T_n^{th}}, 
\end{align}
where $\lambda_n$ is the average incoming packet rate with Poisson distribution, $T_{slt}$ is the duration of a time slot, and the queue can be modeled as M/M/1 queue according to the approximated PDF in \eqref{approximated pdf}. Furthermore, to find the upper bound of $\beta_n$, we know that $P_n^{dly} \le 1$. Thus, the upper bound for Rayleigh and Rician channels are given by: 
\begin{align}
\begin{cases}
    Q_1(b,\beta_n^{Rice}) \ge 1-(1-\lambda_n T_{slt})^{\frac{1}{|F|}},       & \text{Rician;}\\
    \beta_n ^ {Ray} \le \sqrt{-\Omega ln[1-(1-T_{slt} \lambda_n) ^ \frac{1}{|F|}]}         & \text{Rayleigh.}
\end{cases}
\end{align}

\textbf{Buffer Overflow Model.} As mentioned, consider an M/M/1 transmit queue model with a finite buffer size. Thus, upon the arrival of a new packet, it is admitted only if there is a space in the queue. Using queuing theory notion \cite{Gross-2008-Fundamentals} \cite{Roy-2021-Overview}, the probability of exceeding the buffer capacity in a certain state $i$ is determined as follows: 
\begin{align*}
\overline{P_{i,i+1}} = P[X_1+...+X_{i+1} > B_n | X_1+...+X_i \le B_n],
\end{align*}
where $X$ is the packet's length, and $B_n$ is the buffer capacity for $n \in N$. Assuming that the packet's length would be an exponential random variable with parameter $\eta_n$, then the complement of the above expression in which the buffer overflow does not occur can be defined as: 
\begin{align}
P_{i,i+1} = 1 - \overline{P_{i,i+1}} = \frac{1 - \sum_{j=0}^i \frac{(B_n\eta_n)^j}{j!}e^{-B_n\eta_n}}{1 - \sum_{j=0}^{i-1} \frac{(B_n\eta_n)^j}{j!}e^{-B_n\eta_n}}. 
\end{align}
By the Markov chain, the local balance equation is $P_{i+1} = \rho_n(\beta_n) P_{i,i+1}P_{i}$, where $\rho_n(\beta_n) = \frac{\lambda_nT_{slt}}{\mu_n(\beta_n)}$ is the offered load, then $P_i$ can be obtained as:
\begin{align}
\begin{aligned}
P_i = \rho_n^i(\beta_n) \bigg(\prod_{j=0}^{i-1}P_{j,j+1}\bigg)P_0 = 
\\
\rho_n^i(\beta_n) \bigg(1-\sum_{j=0}^{i-1}\frac{(B_n\eta_n)^j}{j!}e^{-B_n\eta_n}\bigg)P_0. 
\end{aligned}
\end{align}
From $\sum_{j=0}^\infty P_j = 1$ and $\rho_n(\beta_n) < 1$, $P_0$ can be calculated as:
\begin{align}
P_0 = \frac{1-\rho_n(\beta_n)}{1-\rho_n(\beta_n) e^{-B_n\eta_n(1-\rho_n(\beta_n))}}. 
\end{align}
Finally, the probability of buffer overflow, which also represents the probability of packet loss, can be approximated as:
\begin{align*}
P_n^{ov}(\beta_n) \approx \sum_{i=0}^\infty \overline{P_{i,i+1}}P_i = \frac{(1-\rho_n(\beta_n))e^{-B_n\eta_n(1-\rho_n(\beta_n))}}{1-\rho_n(\beta_n) e^{-B_n\eta_n(1-\rho_n(\beta_n))}}. 
\end{align*}

\section{Impacts of Interference} \label{impacts of interference}
In this section, we focus on the impact of interference on the main link between the source node and its destination and provide a probability of transmission error due to the high interference from other nodes operating in the same spectrum band. Let us consider that $\gamma_{th}$ represents the threshold for the signal-to-interference-plus-noise ratio (SINR). Transmission error occurs when the SINR is lower than the threshold. Thus, the probability of transmission error is given by: 
\begin{align*}
P_n^{err}(\boldsymbol{\beta}) \triangleq Pb(\gamma_n < \gamma_{th}) = Pb\Biggl(\frac{P_n(\hat{h}_n^f)^2(\Tilde{h}_n^f)^2}{P_{TN}+I_n^f(\boldsymbol{\beta}_{-n})} < \gamma_{th} \Biggl), 
\end{align*}
where $P_n$ is the transmission power of the source node, $P_{TN}  = kTW$ denotes the thermal noise power where $k$ is the Boltzmann’s constant, $T$ represents the temperature, and $W$ would be the bandwidth. Moreover,  $I_n^f(\boldsymbol{\beta}_{-n})$ captures the impact of interference from interferer nodes on the destination node such that $\boldsymbol{\beta}_{-n} \triangleq (\boldsymbol{\beta}_m)_{m \in N \backslash n}$, as defined in \cite{Guan-2016-ToTransmit}. Then, $I_n^f(\boldsymbol{\beta}_{-n})$ can be expressed as: 
\begin{align} \label{interference}
I_n^f(\boldsymbol{\beta}_{-n}) = \sum_{m\in N \backslash n} P_m(\hat{h}_{mn}^f)^2(\Tilde{h}_{mn}^f)^2\alpha_m^f(\beta_m), 
\end{align}
in which, $(\hat{h}_{mn}^f)^2$ and $(\Tilde{h}_{mn}^f)^2$ are the path loss and the square of channel fading coefficient between the interferer nodes and the destination, respectively, Also, $\alpha_m^f(\beta_m)$ equals one if interference node $m$ transmits, and zero otherwise. 

We use a classical stochastic geometry approach to model $I_n^f(\boldsymbol{\beta}_{-n})$ using Gamma distribution function. Ultimately, the probability of transmission error can be defined as:
\begin{align*}
\begin{aligned}
P_n^{err}(\boldsymbol{\beta}) = \int_{\beta_n}^\infty Pb(\Tilde{h}_{n}^f = x) v_n(\frac{P_n (\hat{h}_n^f)^2}{\gamma_{th}} x^2 - P_{TN},\boldsymbol{\beta}_{-n})dx,
\end{aligned}
\end{align*}
where $Pb(\Tilde{h}_{n}^f=x)$ is the PDF of Rayleigh or Rician distribution which is determined by the type of channel and $v_n(x,\boldsymbol{\beta}_{-n})$ is the complementary cumulative distribution function (CCDF) of $I_n^f(\boldsymbol{\beta}_{-n})$ which is given by \cite{Guan-2016-ToTransmit}:
\begin{align*}
\begin{aligned}
v_n(x,\boldsymbol{\beta}_{-n}) \triangleq Pb(I_n^f(\boldsymbol{\beta}_{-n})>x)
= 1-\frac{\varphi(k_n(\boldsymbol{\beta}_{-n}),\frac{x}{\theta_n(\boldsymbol{\beta}_{-n})})}{\Gamma(k_n(\boldsymbol{\beta}_{-n}))}, 
\end{aligned}
\end{align*}
where $\varphi(k_n(\boldsymbol{\beta}_{-n}),\frac{x}{\theta_n(\boldsymbol{\beta}_{-n})}) = \int_0^{\frac{x}{\theta_n(\boldsymbol{\beta}_{-n})}}s^{k_n(\boldsymbol{\beta}_{-n})-1}e^{-s}ds$ denotes the incomplete gamma function and $\Gamma(k_n(\boldsymbol{\beta}_{-n})) = \int_0^{\infty}x^{k_n(\boldsymbol{\beta}_{-n})-1}e^{-x}dx 
$ represents the Gamma function. Also, $k_n(\boldsymbol{\beta}_{-n}) = \frac{(\Tilde{E}[I_n^f(\boldsymbol{\beta}_{-n})])^2}{\Tilde{D}[I_n^f(\boldsymbol{\beta}_{-n})]}$ and $\theta_n(\boldsymbol{\beta}_{-n}) = \frac{\Tilde{D}[I_n^f(\boldsymbol{\beta}_{-n})]}{\Tilde{E}[I_n^f(\boldsymbol{\beta}_{-n})]}$ are the shape and scale parameters where $\Tilde{E}[I_n^f(\boldsymbol{\beta}_{-n})]$ and $\Tilde{D}[I_n^f(\boldsymbol{\beta}_{-n})]$ are the first and second order moments of $I_n^f(\boldsymbol{\beta}_{-n})$, respectively.

\begin{figure}[t]
    \centering
   \includegraphics[width=0.85\linewidth]{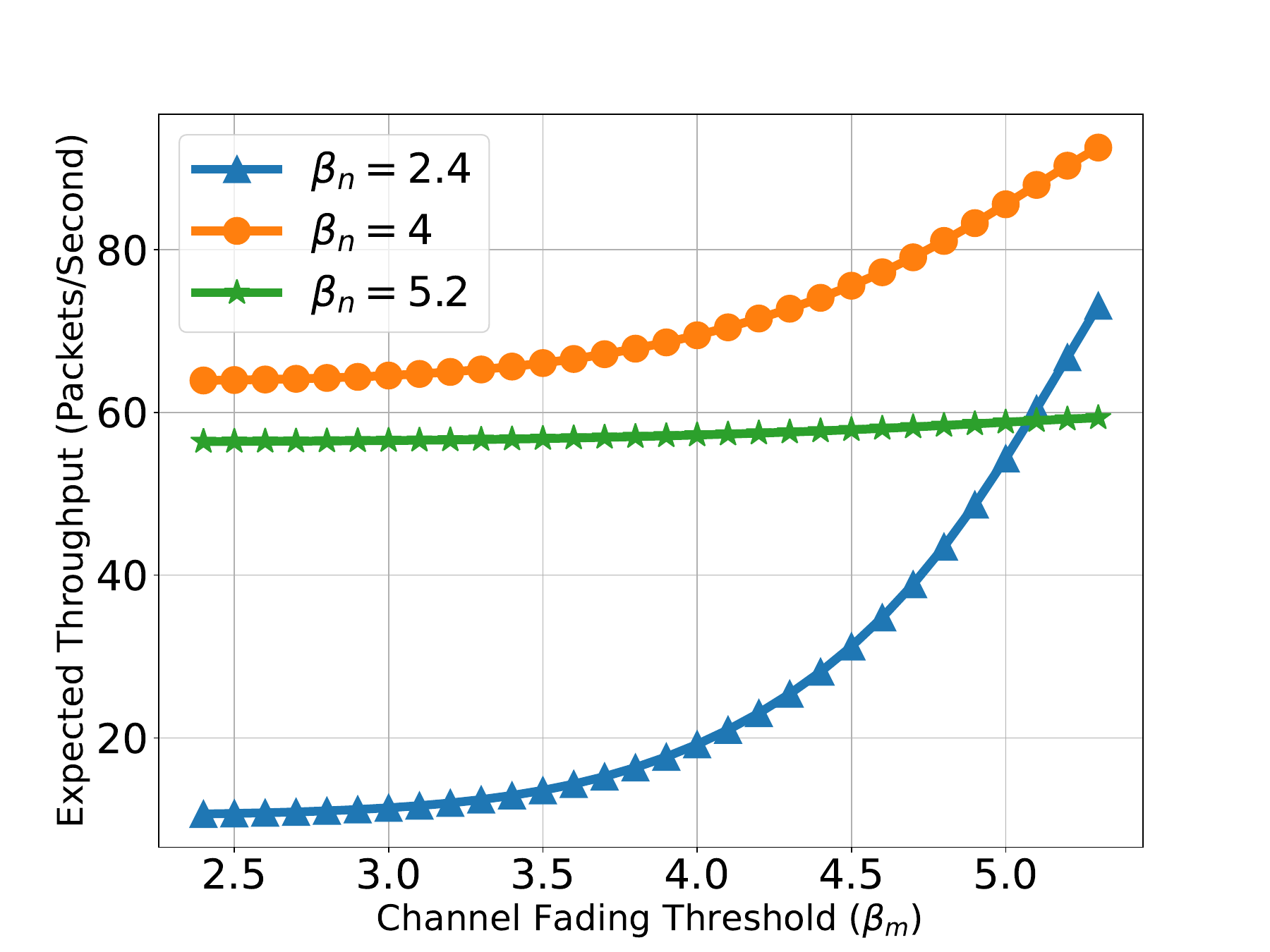}
    \caption{$R_n$ vs. $\boldsymbol{\beta}_m$ by various $\beta_n$}
    \label{Fig. 2}
\end{figure}

\section{Throughput Performance Analysis} \label{throughput performance analysis}
We expressed the probability of packet drop due to exceeding the maximum queuing delay ($P_n^{dly}(\beta_n)$), the probability of buffer overflow due to full buffer ($P_n^{ov}(\beta_n)$), and the probability of transmission error due to low SINR ($P_n^{err}(\boldsymbol{\beta})$). Thus, the probability of overall loss can be expressed as: 
\begin{align}
\begin{aligned}
P_n^{loss}(\boldsymbol{\beta}) = P_n^{ov}(\beta_n) + [1-P_n^{ov}(\beta_n)]P_n^{dly}(\beta_n) + 
\\
[1-P_n^{ov}(\beta_n)][1-P_n^{dly}(\beta_n)]P_n^{err}(\boldsymbol{\beta}). 
\end{aligned}
\end{align}
Since the product of $P_n^{dly}(\beta_n), P_n^{ov}(\beta_n),$ and $P_n^{err}(\boldsymbol{\beta})$ is negligible, the expected throughput can be approximated as:
\begin{align}
\begin{aligned}
R_n(\boldsymbol{\beta}) = \lambda_n [1-P_n^{loss}(\boldsymbol{\beta})] \approx 
\\
\lambda_n [1-P_n^{dly}(\beta_n)-P_n^{ov}(\beta_n)-P_n^{err}(\boldsymbol{\beta})]. 
\end{aligned}
\end{align}

\begin{figure}[t]
    \centering
   \includegraphics[width=0.85\linewidth]{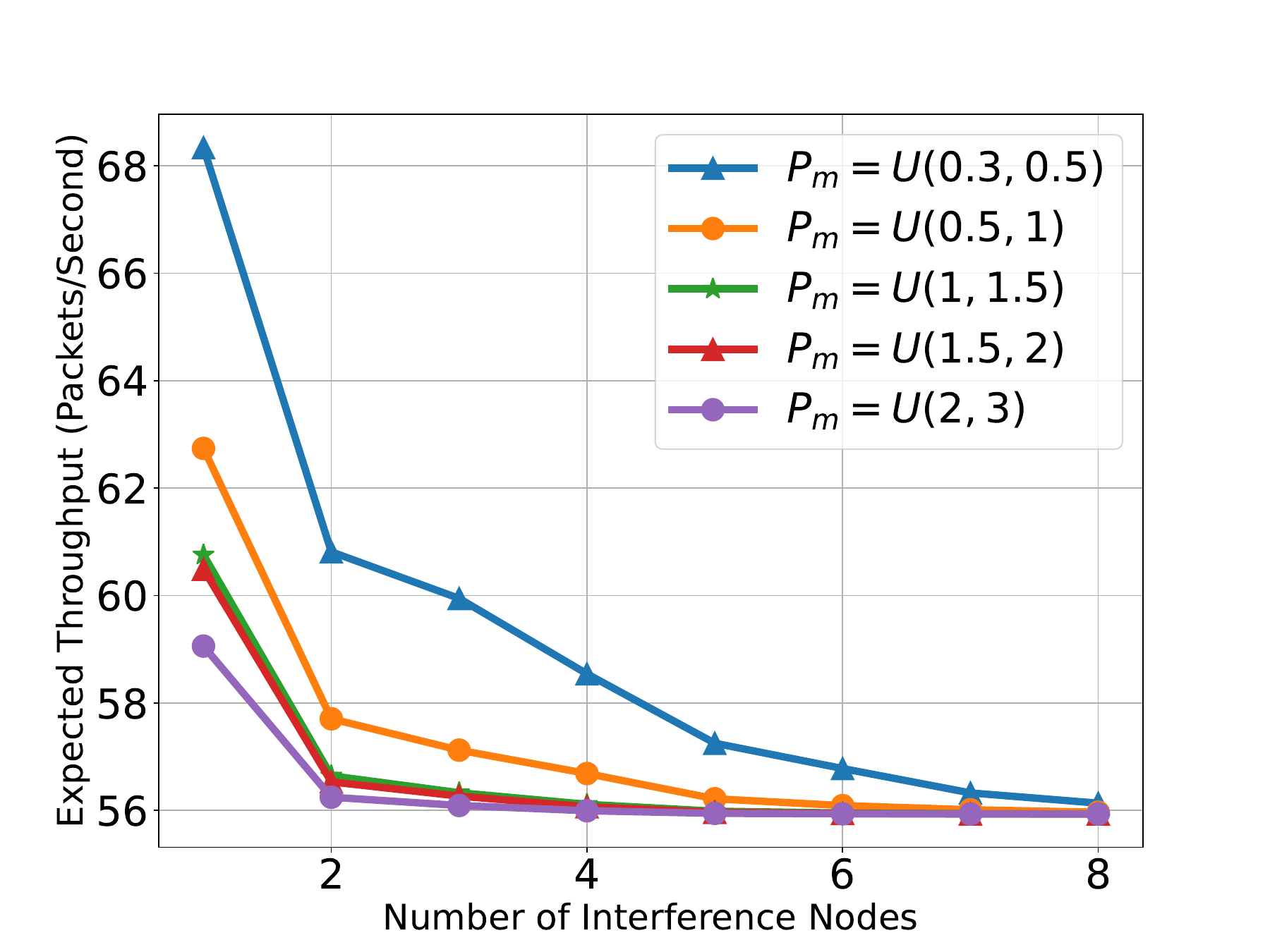}
    \caption{$R_n$ vs. \# of interference nodes by different $P_m$}
    \label{Fig. 3}
\end{figure}


\section{Numerical Results} \label{numerical results}
\textbf{Simulation setup.} In our scenario, we consider one UAV and several ground nodes to see the impact of interference on UAV performance. The main link is always established between one of the ground nodes and the UAV, and other ones are assumed to be interferer nodes in the same region in which all of them are placed according to the Poisson distribution. 

The Rayleigh fading factor is set to $\Omega = 2$, the duration of a time slot is $T_{slt} = 2$ ms, the maximum queuing delay $T_n^{th}$ is uniformly distributed ranging from 30 to 60 ms, the SINR threshold $\gamma_{th}$ equals to 8, the number of frequency channels is $|F| = 15$, reference distance is assumed $d_0 = 20$ m, frequency in the path loss model is set to $f = 900$ MHz, Boltzmann’s constant is $k = 1.38*10^{-23}$, the temperature in the noise model is equal to $T = 290$ K. The communication area is assumed 40 * 40 $m^2$, and the UAV is placed at the altitude of 50 m. The incoming packet rate $\lambda_n$ is uniformly distributed between 60 and 120 with the step size of 20, and the normalized buffer capacity $B_n \eta_n$ is between 50 and 150 with the step size of 25. The transmitter power is considered between 0.5 and 1 watt by uniform distribution. Furthermore, constant parameters in the Rician factor and path loss exponent are set to $k_0=1$, $k_{\frac{\pi}{2}}=15$, $\alpha_0=3.5$, $\alpha_{\frac{\pi}{2}}=2$.

In Fig. \ref{Fig. 2}, $R_n$ is evaluated by different $\boldsymbol{\beta}_m$ and $\beta_n$. The 10 nodes are assumed to establish a Rician channel since we placed the UAV at an altitude of 50 m. Clearly, as $\boldsymbol{\beta}_m$ increases, the interference nodes send fewer packets and enqueue them. Thus, we experience lower interference on the main link, and packets can be easily decoded at the UAV. Also, as $\beta_n$ increases for the source node to the upper bound, we will see different behaviors of $R_n$. The reason is that, as $\beta_n$ starts from the lower values, the impact of $P_n^{err}(\boldsymbol{\beta})$ is more than the $P_n^{dly}(\beta_n)$ and $P_n^{ov}(\beta_n)$ since the source node sends its packets in most of the time even under bad channel conditions. But, as $\beta_n$ gets closer to the upper bound, the source node tries to choose the good channel condition, so more packets will be stored in the queue. Thus, the impact of the packet drop in the queue ($P_n^{dly}(\beta_n)$ and $P_n^{ov}(\beta_n)$) will be more than the packet drop by the transmission error ($P_n^{err}(\boldsymbol{\beta})$).

\begin{figure}
    \centering
   \includegraphics[width=0.85\linewidth]{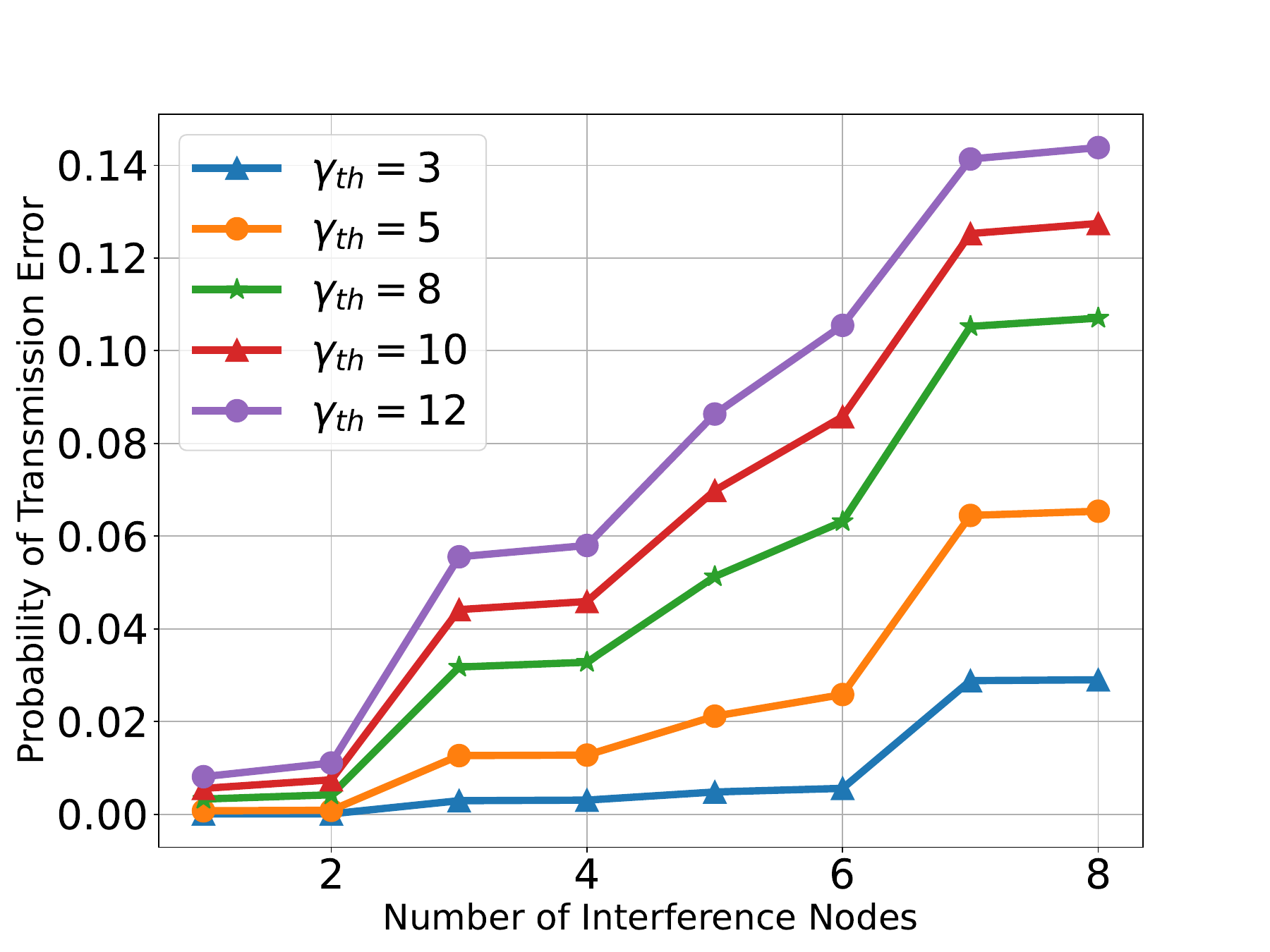}
    \caption{$P_n^{err}$ vs. \# of interference nodes by various $\gamma_{th}$}
    \label{Fig. 4}
\end{figure}

Fig. \ref{Fig. 3} represents the decreasing level of $R_n$ as the number of interference nodes increases. In this scenario, the parameters are $P_n = 0.5$ W, $\beta^{Ray}=1.55$, $\beta^{Rice}=5.1$, and the first two interference nodes are Rician and the other ones are Rayleigh. Clearly, the impact of the first two nodes on $R_n$ is more than the other ones, since Rician interference channels impose more costs on $R_n$. Also, as the transmission power for interference nodes increases, which are uniformly distributed in different ranges, $R_n$ decreases due to a stronger interference.

In Fig. \ref{Fig. 4}, $P_n^{err}$ is represented versus the number of interference nodes where all nodes are established Rician channels. As expected, when the number of interference nodes increases, $P_n^{err}$ increases since the impact of interference would be greater. Also, as $\gamma_{th}$ decreases, $P_n^{err}$ drops since UAV can decode packets correctly, even with a low SINR.

\begin{figure}
    \centering
   \includegraphics[width=0.85\linewidth]{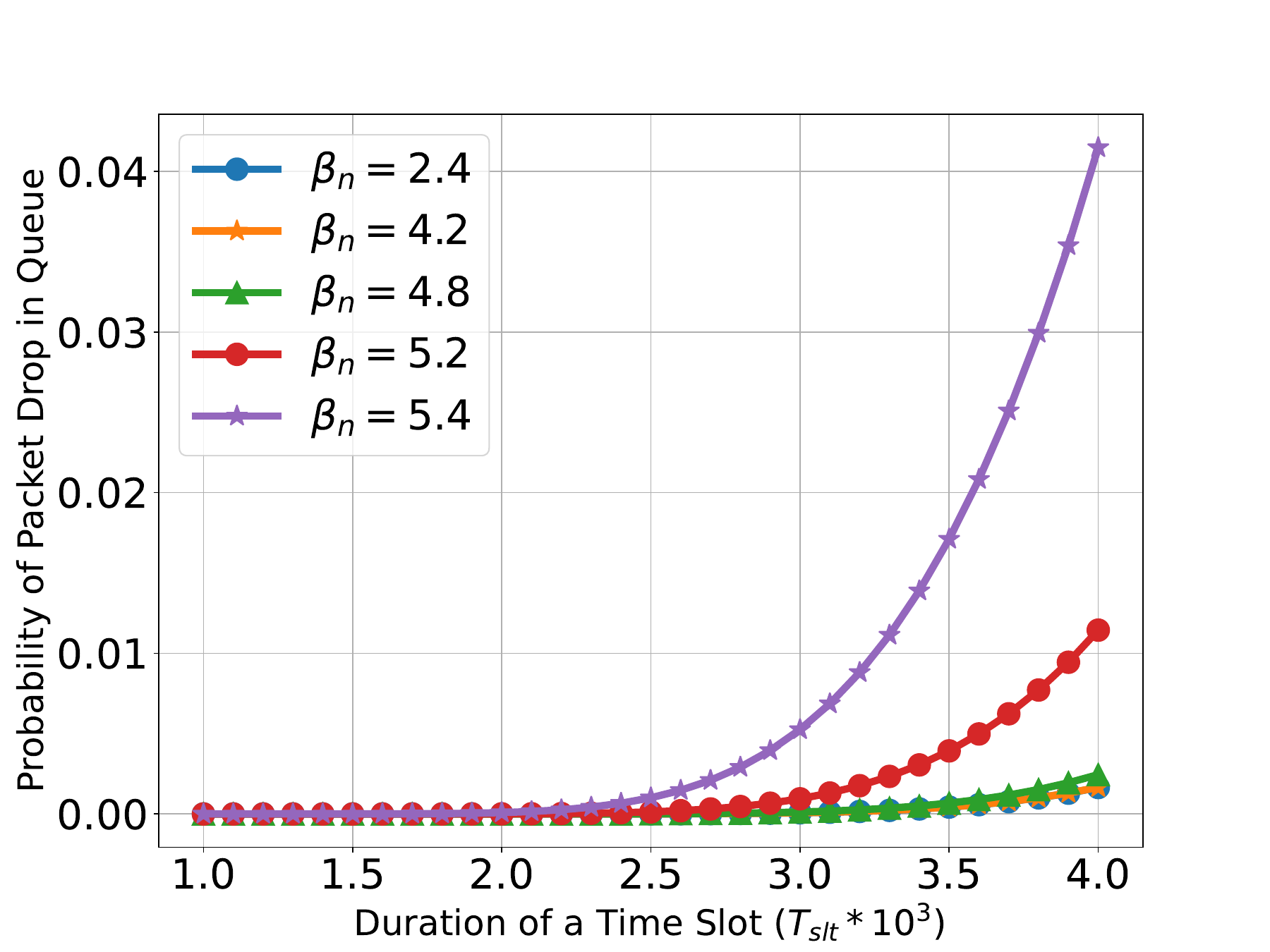}
    \caption{$P_n^{ov} + (1 - P_n^{ov})P_n^{dly}$ vs. $T_{slt}$ by different $\beta_n$}
    \label{Fig. 5}
\end{figure}

Fig. \ref{Fig. 5} shows the probability of packet drop in the queue ($P_n^{ov}(\beta_n) + (1 - P_n^{ov}(\beta_n))P_n^{dly}(\beta_n)$) by changing both $T_{slt}$ and $\beta_n$. The packet drop in the queue means that the packet is dropped due to buffer overflow, or it is not dropped by the buffer overflow, and it is dropped due to exceeding $T_n^{th}$. According to Fig. \ref{Fig. 5}, as $T_{slt}$ grows, the probability of packet drop in the queue increases since packets have less opportunities to be sent. Also, by increasing $\beta_n$ as it gets closer to the upper bound, more packets stay in the queue; thus, $P_n^{dly}$ and, in general, the probability of packet drop in the queue tends to one, but it is not tangible in the lower $\beta_n$.

\section{Conclusion} \label{conclusion}
In this paper, we investigated the expected throughput performance of UAVs when operating in unlicensed spectrum bands. Our framework considers two types of channels, Rayleigh or Rician based on being NLoS or LoS, respectively. 
By considering the impact of interference nodes on UAV, we obtained the packet drop probabilities due to long waiting time in the queue, buffer overflow, and high interferences. Using these expressions, we were able to obtain a general formula to calculate the expected throughput. Furthermore, we analyzed the channel fading threshold ($\beta_n$) for Rayleigh and Rician channels to find the upper bound. In the end, we numerically investigated our scenario by considering the main and some interference links between ground nodes and the UAV to indicate the impact of different parameters.

\section*{Acknowledgment}
The material is based upon work supported by NASA under award No(s) 80NSSC20M0261, and NSF grants 1948511, 1955561, 2212565, and 2323189. Any opinions, findings, and conclusions or recommendations expressed in this material are those of the author(s) and do not necessarily reflect the views of NASA and NSF.

\bibliographystyle{IEEEtran}
\bibliography{ref}

\end{document}